# Multiple Teachers-Meticulous Student: A Domain Adaptive Meta-Knowledge Distillation Model for Medical Image Classification

Shahabedin Nabavi[1], Kian Anvari Hamedani[1], Mohsen Ebrahimi Moghaddam[1], Ahmad Ali Abin[1], Alejandro F. Frangi[2,3,4]

1- Faculty of Computer Science and Engineering, Shahid Beheshti University, Tehran, Iran.
2- Division of Informatics, Imaging and Data Sciences, Schools of Computer Science and Health Sciences, The University of Manchester, Manchester, UK.
3- Medical Imaging Research Center (MIRC), Electrical Engineering and Cardiovascular Sciences Departments, KU Leuven, Leuven, Belgium.
4- Alan Turing Institute, London, UK.

**Corresponding Author:** Mohsen Ebrahimi Moghaddam (PhD)

**Address:** Faculty of Computer Science and Engineering, Shahid Beheshti University, Tehran, Iran.

**Email:** m_moghadam@sbu.ac.ir

**Phone:** +98 912 140 5308

**Conflicts of Interests statement:** There are no conflicts of interests to declare.

**Data availability statement:** The datasets are available at

- York University (YU): http://jtl.lassonde.yorku.ca/software/datasets/
- CMRxRecon: https://cmrxrecon.github.io/
- Automated Cardiac Diagnosis Challenge (ACDC): https://www.creatis.insa-lyon.fr/Challenge/acdc/databases.html
- CMRxMotion: http://cmr.miccai.cloud/
- Movement-related artefacts (MR-ART): https://openneuro.org/datasets/ds004173/versions/1.0.2
- Low-dose Computed Tomography Perceptual Image Quality Assessment Grand Challenge (LDCTIQAG2023): https://zenodo.org/records/7833096#.ZEFywOxBzn5

**Code availability statement:** The codes are available at

https://github.com/kiananvari/MTMS-A-Domain-Adaptive-Meta-Knowledge-Distillation-Model-for-Medical-Image-Classification

**Abstract**

**Background:** Image classification can be considered one of the key pillars of medical image analysis. Deep learning (DL) faces challenges that prevent its practical applications despite the remarkable improvement in medical image classification. The data distribution differences can lead to a drop in the efficiency of DL, known as the domain shift problem. Besides, requiring bulk annotated data for model training, the large size of models, and the privacy-preserving of patients are other challenges of using DL in medical image classification. This study presents a strategy that can address the mentioned issues simultaneously.

**Method:** The proposed domain adaptive model based on knowledge distillation can classify images by receiving limited annotated data of different distributions. The designed multiple teachers-meticulous student model trains a student network that tries to solve the challenges by receiving the parameters of several teacher networks. The proposed model was evaluated using six available datasets of different distributions by defining the respiratory motion artefact detection task.

**Results:** The results of extensive experiments using several datasets show the superiority of the proposed model in addressing the domain shift problem and lack of access to bulk annotated data. Besides, the privacy preservation of patients by receiving only the teacher network parameters instead of the original data and consolidating the knowledge of several DL models into a model with almost similar performance are other advantages of the proposed model.

**Conclusions:** The proposed model can pave the way for practical clinical applications of deep classification methods by achieving the mentioned objectives simultaneously.

**Code Availability:**

https://github.com/kiananvari/MTMS-A-Domain-Adaptive-Meta-Knowledge-Distillation-Model-for-Medical-Image-Classification

## 1. Introduction

Medical image classification refers to the concept of categorizing images based on the appearance of a complication or the recognition of different structures. In the classification process, images with various modalities, i.e., X-ray, CT, MRI, etc., are considered input, and the type of complication or structure under study is regarded as a label/class. Training a learning model to separate the input data based on their class label and achieve the maximum possible efficiency is the ultimate goal of the classification (Badrigilan et al., 2021; Cai et al., 2020). By training the model to a suitable level of generalisability, the trained model can perform well on new unseen data with the same distribution as the training data. However, the performance of learning models drops when faced with data from a different distribution than the training data (Zhou et al., 2022). This issue is a fundamental challenge that exposes learning models to criticism.

Another challenge of medical image classification is the lack of access to bulk annotated datasets (Dhar et al., 2023). Labelling medical imaging data is a time-consuming and tedious process. Meanwhile, for the training of deep learning models, there is a need for a lot of labelled data so that the models achieve the necessary convergence and efficiency. Therefore, due to the lack of large annotated datasets, researchers are looking for the development of methods that can be efficient even with relatively small data (Guan & Liu, 2021; Huisman et al., 2021; Wang et al., 2020; Yu et al., 2022).

Compression of huge and complex deep models or a set of models in a smaller and simpler model with almost similar performance, which can be deployed more easily, is another thing that has recently become a new research trend. Transferring knowledge from a larger model or multiple models to a smaller one is done through a process called knowledge distillation (Gou et al., 2021). In deep neural networks, knowledge is the learned weights or parameters to perform a specific task. In general, the knowledge distillation process requires a distillation approach and a teacher-student architecture (Wang & Yoon, 2021). The knowledge learned by the teacher(s) is transferred through a distillation approach to the student model, which generally has a simpler architecture. After the knowledge distillation process, the student model is expected to perform well on the tasks learned by the teacher or multiple teachers. Besides, there is no need to access the source datasets used in multiple-teacher training to distil knowledge, which can preserve patients' privacy (Chen et al., 2023).

According to the issues raised in the above paragraphs, in this study, we propose a comprehensive model that can cover these issues. The knowledge distillation process in this study is achieved through an architecture including multiple teachers and a meticulous student. The contribution of this study is to propose a novel model that can simultaneously meet the following goals:

- The proposed model can handle the domain shift problem. In other words, the model can perform properly in facing the input dataset with a different distribution than the datasets based on which the teachers were trained.
- Due to the use of meta-learning, the proposed model can achieve proper efficiency even in the conditions of lack of access to considerable annotated data.
- By distilling knowledge, the model takes advantage of the knowledge learned by teachers to learn a new task. Besides, the knowledge of multiple teachers, a set of complex models, is consolidated into a simpler model with almost the same efficiency.
- The proposed model can work, utilising the trained teachers (models' weights) without accessing source datasets. It means there will be no need for the source datasets in the knowledge distillation process, which can lead to the preservation of patients' privacy.

The rest of this article is organised as follows. Section 2 reviews related studies. In section 3, the proposed model, the exploited datasets, and the implementation details are presented. The results of the extensive experiments are given in section 4. Section 5 is dedicated to a discussion of the proposed model and the achieved goals. Finally, the current study is concluded in section 6.

## 2- Related works

Imaging data acquired from the same tissue may differ in distribution due to using different modalities, imaging protocols and parameters, scanners and imaging equipment (Guan & Liu, 2021). This difference in imaging data distribution leads to the domain shift problem that deep learning models cannot achieve the same efficiency in the target domain as those gained in training on the source domain. Domain adaptation approaches can be used to overcome the domain shift problem.

Domain adaptation techniques for medical image analysis can be supervised, semi-supervised and unsupervised (Guan & Liu, 2021). In supervised domain adaptation, mainly a pre-trained model on the source set is fine-tuned with the labelled data of the target set so that the model can be transferred to the target space. The unsupervised domain adaptation techniques use knowledge transfer methods from the annotated source domain to the unlabeled target domain. The main challenges of domain adaptation in medical image analysis include the following:

- Limited access to annotated and bulk datasets in medical applications
- The difference in the medical imaging modalities, which leads to changing the distributions and features of the imaging sets
- The possibility of encountering a situation where there are several source and target sets
- Legal requirements for patients' privacy-preserving issues that require the use of source-free domain adaptation

Unsupervised domain adaptation (Guan & Liu, 2021), meta-learning (Huisman et al., 2021), and few-shot learning (Wang et al., 2020) methods can be leveraged to overcome the challenge of not having access to large annotated datasets, which is the most fundamental challenge in the field. Multi-modality, multi-source/multi-target (Ren et al., 2022) and source-free domain adaptation (Fang et al., 2024) methods are also proposed to face other challenges mentioned. Recently, knowledge distillation has been used in transferring knowledge obtained from one or more source domains to the target domain. Knowledge distillation for domain adaptation could improve the performance of deep learning models in the face of unseen data with different distributions (Nguyen-Meidine et al., 2021).

In some studies, knowledge distillation has been used for unsupervised domain adaptation in medical image segmentation (Liu et al., 2023; X. Wang et al., 2023). In these studies, multi-source domain adaptation methods based on knowledge distillation are proposed for image segmentation. Several studies have also focused on medical image segmentation based on knowledge distillation, but they have not addressed the domain shift problem (C. Fan et al., 2023; You et al., 2023; Zhai et al., 2023). Knowledge distillation has also been leveraged in medical image classification (Tan et al., 2024; Termritthikun et al., 2023; Y. Wang et al., 2023). However, these methods have not concentrated on the domain shift issue. Besides, knowledge distillation has been utilised for medical image classification in long-tailed datasets that contain classes with unbalanced distributions (Elbatel et al., 2023; Zhang et al., 2023). The study of Li et al. (Li et al., 2023) has dealt with medical image enhancement using a source-free unsupervised domain adaptation method based on knowledge distillation. As is clear from the review of the related studies, using knowledge distillation in domain adaptation for medical applications is in its infancy.

## 3- Materials and Methods

The proposed model in this study consists of three main parts, which can be seen in Figure 1. In this section, the problem is formulated first, and then each part of the proposed model, the details of its training process, and the data used are explained.

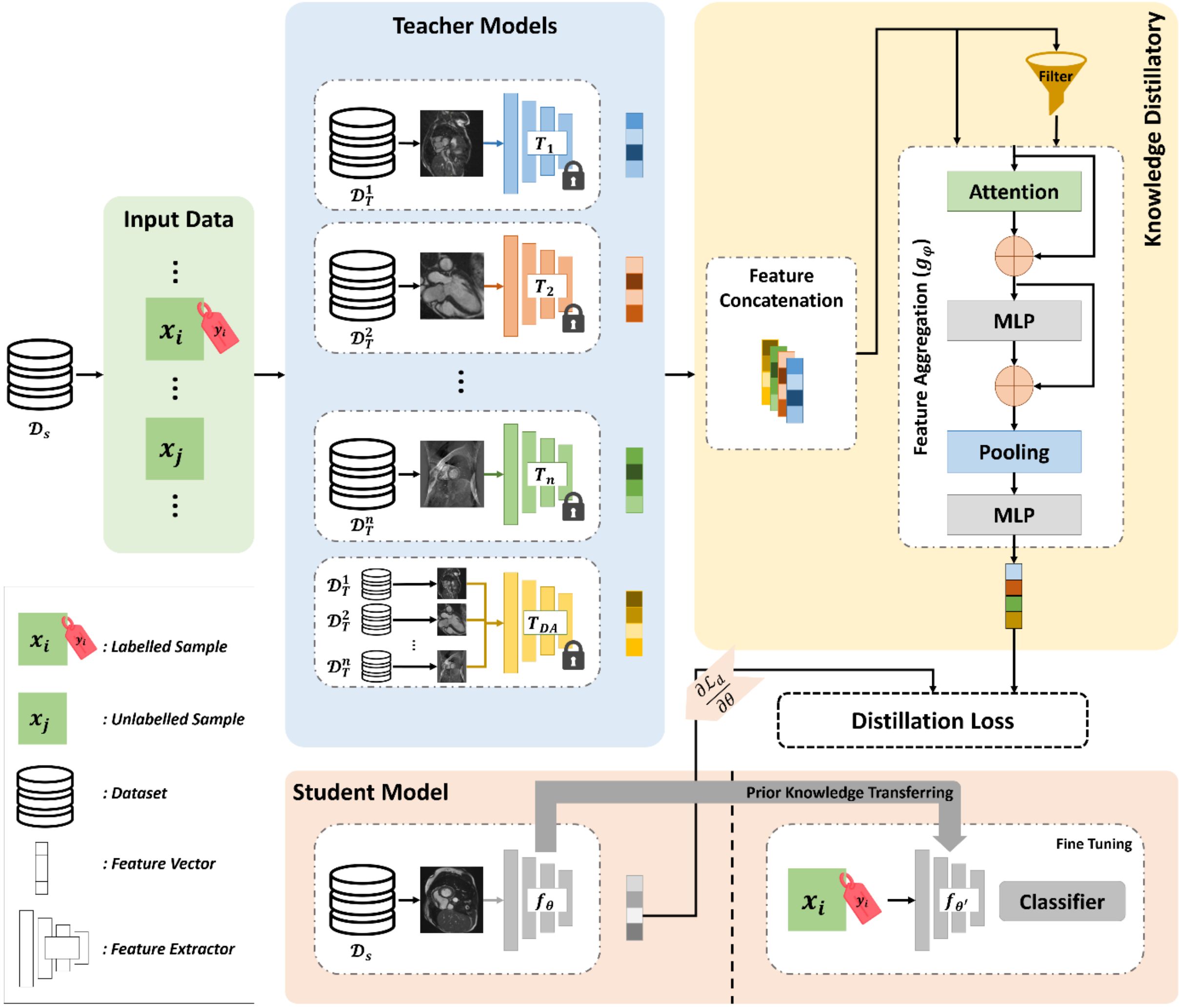

**Figure 1: The overview of the proposed model.**

### 3-1- Problem Formulation

Suppose the task of motion artefact detection in imaging datasets with different distributions, which represent the domain shift between datasets, is on the agenda. Thus, there is a binary-class classification problem. Some datasets are supposed to be used to train multiple teacher models, which are defined in equation (1).

$$\mathcal{D}_T^j = \{(x_i, y_i)\}_{i=1}^{N}, j \in \{1, 2, \dots\} \quad (1)$$

where $x_i$ and $y_i$ represent the image and its corresponding label, respectively. $j$ also specifies the dataset number related to the training of the $j^{th}$ teacher model ($T_j$).

A dataset $\mathcal{D}_S$, in which a tiny subset $\mathcal{D}_S^{Labelled}$ is annotated, is used to train the student model $f_\theta$. Most of the samples ($\mathcal{D}_S^{Unlabelled}$) in $\mathcal{D}_S$ is unlabeled. These datasets are formulated in equations 2 to 4.

$$\mathcal{D}_S = \{\mathcal{D}_S^{Labelled}, \mathcal{D}_S^{Unlabelled}\} \quad (2)$$

$$\mathcal{D}_S^{Labelled} = \{(x_i', y_i')\}_{i=1}^{N'} n \quad (3)$$

$$\mathcal{D}_S^{Unlabelled} = \{x_i''\}_{i=1}^{N''} \quad (4)$$

where $x_i'$ and $y_i'$ indicate the image and its corresponding label, while $x_i''$ is an unlabelled sample. $N'$ and $N''$ show the number of samples of $\mathcal{D}_S^{Labelled}$ and $\mathcal{D}_S^{Unlabelled}$, respectively.

### 3-2- The Proposed Multiple Teachers-Meticulous Student Model

As shown in Figure 1, the proposed model has three main parts:

- The multiple trained teachers
- The knowledge distillatory
- The meticulous student

The proposed multiple teachers-meticulous student model is described in Algorithm 1. The proposed model includes a set of teacher networks that have been trained on their respective datasets. Each dataset used for teacher training has a different distribution than the other datasets and is labelled for the binary classification of the presence/absence of motion artefacts. The knowledge distillatory is supposed to distil the knowledge learned by the teachers in the learning of the student network. Finally, in addition to learning the knowledge of teachers and extracting knowledge that is more efficient for its goals, the meticulous student model learns the desired task in a new dataset, which has a large amount of unlabelled data and a different distribution. In the following, each part is explained in detail.

**Algorithm 1** The Proposed Multiple Teachers-Meticulous Student Model

**Require:**

$\triangleright$ $\mathcal{T}_j$ is a teacher network trained on $\mathcal{D}_T^j = \{(x_i, y_i)\}_{i=1}^N$, where $j \in \{1, 2, ...\}$

$\triangleright$ $\mathcal{T}_{DA}$ is a domain adaptation teacher network trained on all $\mathcal{D}_T^j$

$\triangleright$ $\mathcal{D}_S = \{\mathcal{D}_s^{Labelled}, \mathcal{D}_s^{Unlabelled}\}$, where $\mathcal{D}_s^{Labelled} = \{(x_i^{'}, y_i^{'})\}_{i=1}^{N^{'}}$, $x_i^{'}$ is an annotated image, and $\mathcal{D}_s^{Unlabelled} = \{x_i^{''}\}_{i=1}^{N^{''}}$, $x_i^{''}$ is an unlabelled image

$\triangleright$ $\theta$ : Initial parameters of the student network

$\triangleright$ $c_\theta$ : Initial parameters of the classifier network for the student

$\triangleright$ $\varphi$ : Initial parameters of the feature aggregator network

$\triangleright$ $c_\varphi$ : Initial parameters of the classifier network for the feature aggregator

**Ensure:**

$\triangleright$ $\theta^{'}$ : Final parameters of meticulous student network for classification task

$\triangleright$ $c_\theta^{'}$ : Final parameters of classifier network for classification task

$\triangleright$ Prior Meticulous Student Model Training :

**for** $iteration = 1, 2, ...$ **do**
  **for** All mini-batches from $\mathcal{D}_S$ **do**
    **for** All $z_i \in$ mini-batch **do**
      **for** $j = 1, 2, ..., N+1$ **do**
        $e_{ij} \leftarrow T_j(z_i)$
      **end for**
      **if** $z_i \in \mathcal{D}_s^{Unlabelled}$ **then**
        $e_i^{KD} \leftarrow g_\varphi(concat(e_{ij}))$
      **else**
        $e_i^{KD} \leftarrow g_\varphi(F(concat(e_{ij})))$
      **end if**
      $e_i^{ST} \leftarrow f_\theta(z_i)$
      $\mathcal{L}_i^{Dist}(e_i^{KD}, e_i^{ST}) = \|e_i^{KD} - e_i^{ST}\|_2$
    **end for**
    $\mathcal{L}_b^{Dist} = \sum_{i=1}^{batch_size} \mathcal{L}_i^{Dist}(e_i^{KD}, e_i^{ST})$
    $\theta \leftarrow$ Adam_Optimizer $(\mathcal{L}_b^{Dist}, \theta)$
  **end for**
  **for** $(x_i^{'}, y_i^{'}) \in D_s^{Labelled}$ **do**
    $\hat{y}_i^\theta \leftarrow f_{\theta, c_\theta}(x_i^{'})$
    $\mathcal{L}_{CE}(\hat{y}_i^\theta, y_i^{'}) = -\frac{1}{N'} \sum_{i=1}^{N^{'}} (y_i^{'}. \log(\hat{y}_i^\theta) + (1 - y_i^{'}). \log(1 - \hat{y}_i^\theta))$
    $\hat{y}_i^\varphi \leftarrow g_{\varphi, c_\varphi}(F(concat(T_j(x_i^{'}))))$
    $\mathcal{L}_{CE}(\hat{y}_i^\varphi, y_i^{'}) = -\frac{1}{N'} \sum_{i=1}^{N^{'}} (y_i^{'}. \log(\hat{y}_i^\varphi) + (1 - y_i^{'}). \log(1 - \hat{y}_i^\varphi))$
    $\theta \leftarrow$ Adam_Optimizer $(\mathcal{L}_{CE}(\hat{y}_i^\theta, y_i^{'}), \theta)$
    $c_\theta \leftarrow$ Adam_Optimizer $(\mathcal{L}_{CE}(\hat{y}_i^\theta, y_i^{'}), c_\theta)$
    $\varphi \leftarrow$ Adam_Optimizer $(\mathcal{L}_{CE}(\hat{y}_i^\varphi, y_i^{'}), \varphi)$
    $c_\varphi \leftarrow$ Adam_Optimizer $(\mathcal{L}_{CE}(\hat{y}_i^\varphi, y_i^{'}), c_\varphi)$
  **end for**
**end for**

$\triangleright$ Prior Model Fine–tuning :

**for** $iteration = 1, 2, ...$ **do**
  **for** $(x_i^{'}, y_i^{'}) \in D_s^{Labelled}$ **do**
    $\hat{y}_i^\theta \leftarrow f_{\theta, c_\theta}(x_i^{'})$
    $\mathcal{L}_{fine_tune}(\hat{y}_i^\theta, y_i^{'}) = -\frac{1}{N'} \sum_{i=1}^{N^{'}} (y_i^{'}. \log(\hat{y}_i^\theta) + (1 - y_i^{'}). \log(1 - \hat{y}_i^\theta))$
    $\theta^{'} \leftarrow$ Adam_Optimizer $(\mathcal{L}_{fine_tune}(\hat{y}_i^\theta, y_i^{'}), \theta)$
    $c_\theta^{'} \leftarrow$ Adam_Optimizer $(\mathcal{L}_{fine_tune}(\hat{y}_i^\theta, y_i^{'}), c_\theta)$
  **end for**
**end for**

**return** $\theta^{'}, c_\theta^{'}$ $\triangleright$ Final Parameters of the Meticulous Student

### 3-2-1- The Multiple Trained Teachers

The teachers, each trained on its dedicated dataset in a supervised manner, are received to form a multiple-teacher model. These trained teachers should generate an embedding by receiving each sample $z_i \in \mathcal{D}_s$. Thus, each teacher model is defined as equation 5:

$$T_j: z_i \rightarrow e_{ij}, \quad i \in \{1, 2, \dots, N' + N''\}, j \in \{1, 2, \dots\} \tag{5}$$

where $z_i$ is the $i^{th}$ sample from dataset $\mathcal{D}_s$ and can be $x_i'$ (labelled) or $x_i''$ (unlabelled). $e_{ij}$ also represents the embedding obtained from the $j^{th}$ teacher for the $i^{th}$ sample of $\mathcal{D}_s$.

The teacher networks are each trained on a dataset with a different distribution than other datasets, and their weights are unchanged during the training of the proposed model. In addition to teacher networks $T_1$ to $T_n$, a teacher $T_{DA}$ is trained on all $\mathcal{D}_T^j$ datasets to extract domain-independent features. This teacher network, whose weights are also intact during the training of the proposed model, has the task of extracting imaging features between these datasets when domain shift between datasets $\mathcal{D}_T^j$ exists. Teacher $T_{DA}$ works based on a domain adaptation approach. Therefore, it is expected that each teacher from $T_1$ to $T_n$ can extract suitable features in their respective domains, and $T_{DA}$ also obtains common features between different domains.

### 3-2-2- The Knowledge Distillatory

The knowledge distillation process is carried out by the knowledge distillatory part, which uses the concatenation of the embeddings obtained from teachers' networks and then combines them to form a new embedding ($e_i^{KD}$). In the case that the input data to the proposed model is $z_i \in \mathcal{D}_S^{Unlabelled}$, the knowledge distillation network, which has a transformer structure, uses the concatenation of all embeddings $e_{ij}$ obtained from all teachers. Meanwhile, when the input of the proposed model is $z_i \in \mathcal{D}_S^{Labelled}$, because $z_i$ has a label, embeddings $e_{ij}$, whose teacher networks correctly predict the label of $z_i$, can be selected, and other embeddings are masked. This masking of embeddings is done by the filter part ($F$) of the knowledge distillatory and results in the focus of knowledge distillation on more valuable features. The concatenated embeddings for $z_i \in \mathcal{D}_S^{Unlabelled}$ and the masked concatenated embeddings for $z_i \in \mathcal{D}_S^{Labelled}$ are fed as input to the feature aggregator network ($g_\varphi$). This network aggregates input embeddings and creates a new embedding ($e_i^{KD}$) that is used to train the student's network ($f_\theta$). The feature aggregator network is formulated in equation 6.

$$g_\varphi: \begin{cases} Concat(e_{ij}) \rightarrow e_i^{KD} & if\ z_i \in \mathcal{D}_S^{Unlabelled} \\ F\left(Concat(e_{ij})\right) \rightarrow e_i^{KD} & if\ z_i \in \mathcal{D}_S^{Labelled} \end{cases} \tag{6}$$

### 3-2-3- The Meticulous Student

We have called the student in the proposed model meticulous because it uses the minimal annotated data available to improve the learning process. Using annotated data to mask the embeddings that the teachers who generated them failed to recognise the true labels makes the feature aggregator network learn more suitable embeddings. Accordingly, the meticulous student tries to use the knowledge of teachers who are more useful for its purposes than others through embeddings that have made a correct prediction.

The student network is supposed to learn the desired task by receiving data from dataset $\mathcal{D}_s$, which is mostly unlabelled. In addition to learning the classification task on dataset $\mathcal{D}_s$, it is expected that the student network can perform well on other datasets ($\mathcal{D}_T^j$) by distilling teachers' knowledge. Thus, with the proposed model, the student network should be obtained, which can overcome the domain shift problem, achieve a suitable performance on the target dataset ($\mathcal{D}_s$) using a tiny volume of annotated data ($\mathcal{D}_s^{Labelled}$), and gather the knowledge obtained from teachers and training on the target dataset in a minimal network. For this purpose, the student network is first defined as equation 7.

$$f_\theta: z_i \to e_i^{ST}, z_i \in \mathcal{D}_s \tag{7}$$

where the student model ($f_\theta$) produces an embedding $e_i^{ST}$ for the $i^{th}$ input $z_i \in \mathcal{D}_s$.

The student network should try to make its output embedding ($e_i^{ST}$) similar to the embedding ($e_i^{KD}$) obtained from $g_\varphi$. The distillation loss function is considered as equation 8 to achieve this goal.

$$\mathcal{L}_i^{Dist} = \begin{cases} \left\| g_\varphi\left(concat\left(T_j(z_i)\right)\right) - f_\theta(z_i)\right\|_2 & if\ z_i \in \mathcal{D}_s^{Unlabelled} \\ \left\| g_\varphi\left(F\left(concat\left(T_j(z_i)\right)\right)\right) - f_\theta(z_i)\right\|_2 & if\ z_i \in \mathcal{D}_s^{Labelled} \end{cases} \tag{8}$$

where $\mathcal{L}_i^{Dist}$ represents the calculated distillation loss for the $i^{th}$ input, which can be labelled or unlabelled.

By choosing a mini-batch of $\mathcal{D}_s$ as the support set, the distillation loss function is calculated using equation 9.

$$\mathcal{L}_b^{Dist} = \sum_{i=1}^{batch_size} \mathcal{L}_i^{Dist}(\, e_i^{KD}, e_i^{ST}) \tag{9}$$

where $\mathcal{L}_b^{Dist}$ is the distillation loss obtained for the selected mini-batch $b$.

Given $\mathcal{L}_b^{Dist}$, the parameters of the student model are updated using Adam optimiser (Kingma & Ba, 2015). Equation 10 shows how to update $f_\theta$ parameters. This update of $f_\theta$ parameters is repeated for every mini-batch.

$$\theta \leftarrow Adam_optimizer\ (\mathcal{L}_b^{Dist}, \theta) \tag{10}$$

At the end of each epoch, the student model has been fully trained using all the training data of $\mathcal{D}_s$ once, and now the $\mathcal{D}_s^{Labelled}$ is used as the query set to update $g_\varphi$ and $f_\theta$. For this purpose, a classifier $f_c$ is added to the end of the feature aggregator $g_{\varphi}$ and the student network $f_{\theta}$ so that the loss function can be calculated using the annotated data according to Equation 11.

$$\mathcal{L}_{CE}(\hat{y}',y') = -\frac{1}{N'}\sum_{i=1}^{N'}(y_i' \,.\log(\hat{y}_i') + (1-y_i').\log(1-\hat{y}_i')) \tag{11}$$

where $y'$ is the actual label of the input data and $\hat{y}'$ is the value obtained from $f_c$. Besides, $N'$ indicates the size of $\mathcal{D}_S^{Labelled}$.

Then, the $g_{\varphi}$, $f_{\theta}$ and $f_c$ parameters are updated using Adam optimiser with the $\mathcal{L}_{CE}(\hat{y}',y')$ obtained at the end of each epoch. Equation 12 expresses these parameter updates.

$$< \varphi;\ \theta;\ c > \leftarrow Adam_optimizer\ (\mathcal{L}_{CE}(\hat{y}',y'), < \varphi;\ \theta;\ c >) \tag{12}$$

After performing the above steps and completing all epochs, $f_{\theta}$ is obtained as the prior model, which contains meta-knowledge distilled from teachers and learned from $\mathcal{D}_S$. Now, this prior model must be fine-tuned in several epochs to achieve proper performance. Fine-tuning of $f_{\theta}$ and $f_c$ is done on $\mathcal{D}_S^{Labelled}$, and Equation 13 is leveraged as a loss function. Then, the parameters are updated using Adam optimiser according to Equation 14, and the final model $f_{\theta'}$ is obtained.

$$\mathcal{L}_{fine_tune}(\hat{y}',y') = -\frac{1}{N'}\sum_{i=1}^{N'}(y_i' \,.\log(\hat{y}_i') + (1-y_i').\log(1-\hat{y}_i')) \tag{13}$$

$$< \theta';\ c' > \leftarrow Adam_optimizer\ (\mathcal{L}_{fine_tune}(\hat{y}',y'), < \theta;\ c >) \tag{14}$$

**3-3- Dataset Description and Preprocessing**

Several available datasets were selected to evaluate the proposed model in various settings. The selected datasets have differences in data distribution or domain shift and are chosen from different modalities. The dataset of cardiovascular magnetic resonance (CMR) images of the York University (YU) (Andreopoulos & Tsotsos, 2008) acquired from 33 under-18-years of age subjects contains 7980 long- and short-axis images. For each subject, the long-axis view includes between 8 and 15 slices, and the short-axis view has 20 frames. The CMRxRecon dataset related to the MICCAI challenge 2023 contains short- and long-axis raw data of 300 healthy subjects. For the short-axis cine view, 5 to 10 slices and one slice each for the long-axis views were acquired using a Siemens 3T scanner in 12 to 25 phases in the cardiac cycle. ACDC dataset containing short-axis images of 150 patients is another dataset used in the current study. To know the details of this dataset, refer to (Bernard et al., 2018). The CMRxMotion dataset released for the MICCAI challenge 2022 includes labelled CMR images of 40 healthy volunteers. The imaging data of this dataset are in four levels of respiratory artefact labelled by radiologists. One of the goals of releasing this dataset is CMR image quality assessment from the perspective of respiratory motion artefact. Movement-related artefacts (MR-ART) dataset includes T1-weighted head MR images of 148 healthy individuals. The subjects' heads were imaged in three modes: without movement, with slight and immoderate head motions, so the images have three labels of motion artefacts. For more details on this dataset, refer to (Nárai et al., 2022). Low-dose computed tomography perceptual image quality assessment grand challenge (LDCTIQAG2023) dataset (Wonkyeong Lee, 2023) has been released for the MICCAI challenge 2023. The image quality assessment problem based on the radiologists' opinion score has been targeted in this dataset. For cross-modality evaluations, this dataset has also been used in the current study.

The proposed model in the current study provides a general strategy for medical image classification in the presence of domain shift, lack of access to large annotated data, and lack of access to some datasets due to preserving patients' privacy. To check the effectiveness of the proposed model, we considered the task of respiratory motion artefact detection in medical images, which is one of the most important topics in medical image analysis. The datasets used are various modalities such as short- and long-axis CMR images, brain MRI and CT. Some of these datasets, including CMRxMotion, MR-ART and LDCTIQAG2023, have real respiratory artefacts that occurred during imaging, while in others, these artefacts were artificially added to the images. To add respiratory artefacts to YU, CMRxRecon and ACDC datasets that include CMR images, the k-space manipulation method was used, like previous studies (Lorch et al., 2017; Nabavi et al., 2021; Oksuz et al., 2019). The reference images are subjected to a one-dimensional shift and then transferred to the corresponding k-space using the Fourier transform. Combining the k-space lines from the reference image and the corresponding translated one based on a sinusoidal-pattern results in the degraded k-space generation with realistic respiratory artefacts.

5,000 good-quality images were selected from the YU, CMRxRecon(SAX), CMRxRecon(LAX), and ACDC datasets to conduct experiments, then 5,000 corrupted images were generated using other good-quality images from these datasets and the k-space manipulation method. Thus, 10,000 images are considered for each of the mentioned datasets, and to balance the classes, each class contains 5,000 images. Besides, for CMRxMotion, MR-ART and LDCTIQAG2023 datasets, each class has 755, 5000, and 118 images, respectively.

**3-4- Implementation Details**

Teacher models ($T_j$) use ResNet18 (He et al., 2016) as a feature extraction network, which is supplemented with two fully connected layers as a classifier, including 512 and 1 artificial neurons. The domain-adaptation teacher ($T_{DA}$), which is supposed to obtain domain-independent features, consists of ResNet18 as a feature extractor and two classifiers to identify labels and domains. The label classifier, which has the task of identifying the classes of the presence or absence of respiratory artefacts, consists of fully connected layers, including 512 and 1 neurons. The domain classifier is supposed to be unable to recognise the domain label of each image, which means that its error in identifying which dataset the input image is from increases. The structure of (512 neurons+ 100 neurons+ batch-normalisation (BN)+ Relu+ 3 neurons) is used for the domain classifier network, which uses the LogSoftMax activation function in the last layer. The structure of the domain-adaptation teacher is inspired by the (Ganin & Lempitsky, 2015) model. The training of teacher models has been done in 20 epochs and with batch-size=16 in a supervised manner. The loss function and the optimisation method used are cross-entropy and Adam with a learning_rate=0.001, respectively.

The meticulous student model is trained in two stages: prior and fine-tuning. In the prior stage, $f_{\theta}$, which is a ResNet18, followed by $c_{\theta}$ containing layers with 512 and 1 neurons with sigmoid activation function, tries to bring the embedding obtained from the feature extractor close to the embedding obtained from $g_{\varphi}$. The distillation loss (Equation 8) is used to update the weights of the meticulous student model in the prior stage. The $g_{\varphi}$ model is based on the transformer structure and is shown in the knowledge distillatory section of Figure 1. A classifier $c_{\varphi}$ containing 512 neurons followed by 1 neuron is used to train $g_{\varphi}$, which uses binary cross-entropy as a loss function. In the fine-tuning stage, the classifier $c'_{\theta}$ is leveraged, including layers of 512 and 1 neurons. The loss function at the fine-tuning stage is the binary cross-entropy. The prior and fine-tuning stages use Adam optimiser with learning_rate of 0.001 and 0.0001, respectively. The training of the meticulous student model is done in the prior stage in 40 epochs with a batch size of 128 and in the fine-tuning stage in 20 epochs with a batch size of 8.

All the codes related to the current study have been developed in Python programming language using Pytorch and torchvision packages (Paszke et al., 2019). Matplotlib, Seaborn and sklearn packages are also leveraged for the visualisation of results. All experiments were performed using a computer system equipped with NVIDIA GeForce GTX TITAN X GPU and 16GB of RAM.

### 3-5- Experimental Setup

Extensive experiments were designed and conducted to prove the effectiveness of the proposed model. These experiments are repeated by changing the size of $\mathcal{D}_S^{Labelled}$ to clarify the improvement achieved by the proposed method. These experiments are as follows:

- Conducting tests to obtain multiple teacher models with appropriate performance.
- Training and testing the meticulous student model using various target datasets with different distributions under the conditions of using $\mathcal{D}_S^{Labelled}$ with sizes of 32 and 64 samples (16 and 32 samples from each class).
- Investigations related to the change in the number of teachers and its effect on the results of the meticulous student model.
- Conducting ablation studies to check the effectiveness of different elements of the proposed model including the $T_{DA}$ and the filter $F$.
- Comparisons to evaluate the efficiency of the proposed model compared to the related studies.

The aforementioned tests were performed in 5-fold cross-validation to increase the reliability of the obtained results. By changing the size of $\mathcal{D}_S^{Labelled}$, the generalisability of the proposed model was checked. The results of the experiments are also supported with appropriate diagrams so that the significance of the proposed model can be seen intuitively. Accuracy (ACC), precision (PR), recall (RE), and F1-score metrics (Powers, 2020) have been used for quantitative evaluation. Besides, datasets have been split into 80% for training and validation and 20% for test sets.

## 4- Results

### 4-1- Analysis of the Proposed Model

An experiment was conducted with the definition of four teacher networks to evaluate the performance of the proposed model. In this experiment, using YU, CMRxRecon(LAX) and CMRxRecon(SAX) datasets, three networks were trained in a supervised manner to be used as teachers in the proposed model. Also, the teacher model $T_{DA}$ was trained using all three of these datasets to extract domain-independent features. The multiple-teachers part of the proposed model is prepared with these four teachers, and we named each teacher based on the dataset used in its training. The results obtained from testing the teacher models are shown in Table 1.

**Table 1: Results of supervised training of teacher models.**

| Teacher \ Metric | Accuracy | Precision | Recall | F1-score |
|---|---|---|---|---|
| $T_{YU}$ | 99.77 ± 0.14 | 100.0 ± 0.0 | 99.55 ± 0.28 | 99.77 ± 0.14 |
| $T_{CMRxRecon(LAX)}$ | 99.83 ± 0.04 | 100.0 ± 0.0 | 99.66 ± 0.09 | 99.83 ± 0.04 |
| $T_{CMRxRecon(SAX)}$ | 99.84 ± 0.05 | 100.0 ± 0.0 | 99.68 ± 0.11 | 99.84 ± 0.05 |
| $T_{DA}$ | 99.81 ± 0.20 | 100.0 ± 0.0 | 99.63 ± 0.39 | 99.81 ± 0.20 |

Having the trained teacher models, other datasets were tested on these models to investigate the domain shift between different datasets. The results of these tests are listed in Table 2. As can be seen, for example, although the ACDC and CMRxMotion datasets contain CMR images like the YU, CMRxRecon(LAX) and CMRxRecon(SAX) datasets, the performance of the teacher models has dropped in the analysis of these datasets. This problem is caused by the difference in domain or distribution between different datasets. Undoubtedly, the decrease in the performance of the teacher models on the other datasets is also higher due to incompatible contents and modalities.

**Table 2: The results of testing the target datasets on the trained teacher networks. The numbers in parentheses indicate the standard deviation.**

| | $T_{YU}$ | | | | $T_{CMRxRecon(LAX)}$ | | | | $T_{CMRxRecon(SAX)}$ | | | | $T_{DA}$ | | | |
|---|---|---|---|---|---|---|---|---|---|---|---|---|---|---|---|---|
| | ACC | PR | RE | F1 | ACC | PR | RE | F1 | ACC | PR | RE | F1 | ACC | PR | RE | F1 |
| ACDC | 70.63 (8.27) | 74.50 (9.08) | 70.42 (9.23) | 67.75 (11.23) | 64.72 (6.60) | 61.37 (6.30) | 89.98 (10.23) | 71.22 (8.29) | 59.57 (6.60) | 57.13 (4.87) | 89.23 (12.49) | 68.69 (5.71) | 82.70 (9.34) | 77.65 (11.2) | 99.14 (1.37) | 86.05 ( 7.29) |
| CMRxMotion | 57.68 (3.59) | 64.13 (5.10) | 44.13 (22.19) | 48.82 (16.31) | 59.47 (2.05) | 58.48 (3.45) | 77.29 (12.95) | 65.89 (3.15) | 57.55 (4.94) | 56.12 (4.05) | 85.03 (10.22) | 67.24 (2.60) | 64.50 (1.99) | 69.96 (5.87) | 57.59 (11.58) | 62.31 (5.66) |
| MR-ART | 56.15 (3.25) | 55.44 (32.66) | 24.80 (17.57) | 32.53 (19.91) | 50.50 (0.17) | 40.00 (8.94) | 0.02 (0.04) | 0.04 (0.09) | 51.4 (1.71) | 32.61 (23.21) | 3.83 (8.39) | 5.92 (12.90) | 52.05 (1.13) | 79.60 (22.12) | 16.95 (2.32) | 3.25 (4.43) |

| | | | | | | | | | | | | | | | |
|---|---|---|---|---|---|---|---|---|---|---|---|---|---|---|---|
| LDCTIQAG2023 | 41.30 (0.00) | 10.0 (22.37) | 0.74 (1.66) | 1.38 (3.08) | 59.13 (20.99) | 53.07 (34.48) | 57.78 (33.34) | 53.92 (36.89) | 43.48 (4.86) | 25.56 (36.33) | 6.67 (11.23) | 10.36 (16.90) | 58.24 (9.95) | 30.26 (21.22) | 21.21 (32.21) | 23.91 (33.99) |

In addition to the results obtained in Table 2, which confirm the domain shift between the studied datasets, Figure 2 also shows the distribution of samples in each dataset. The distribution of these datasets shows the diversity of the data and, thus, can support the results of Table 2. Overall, the results show that by training a deep learning model on one or more of these datasets, one cannot expect proper performance on unobserved data for the model.

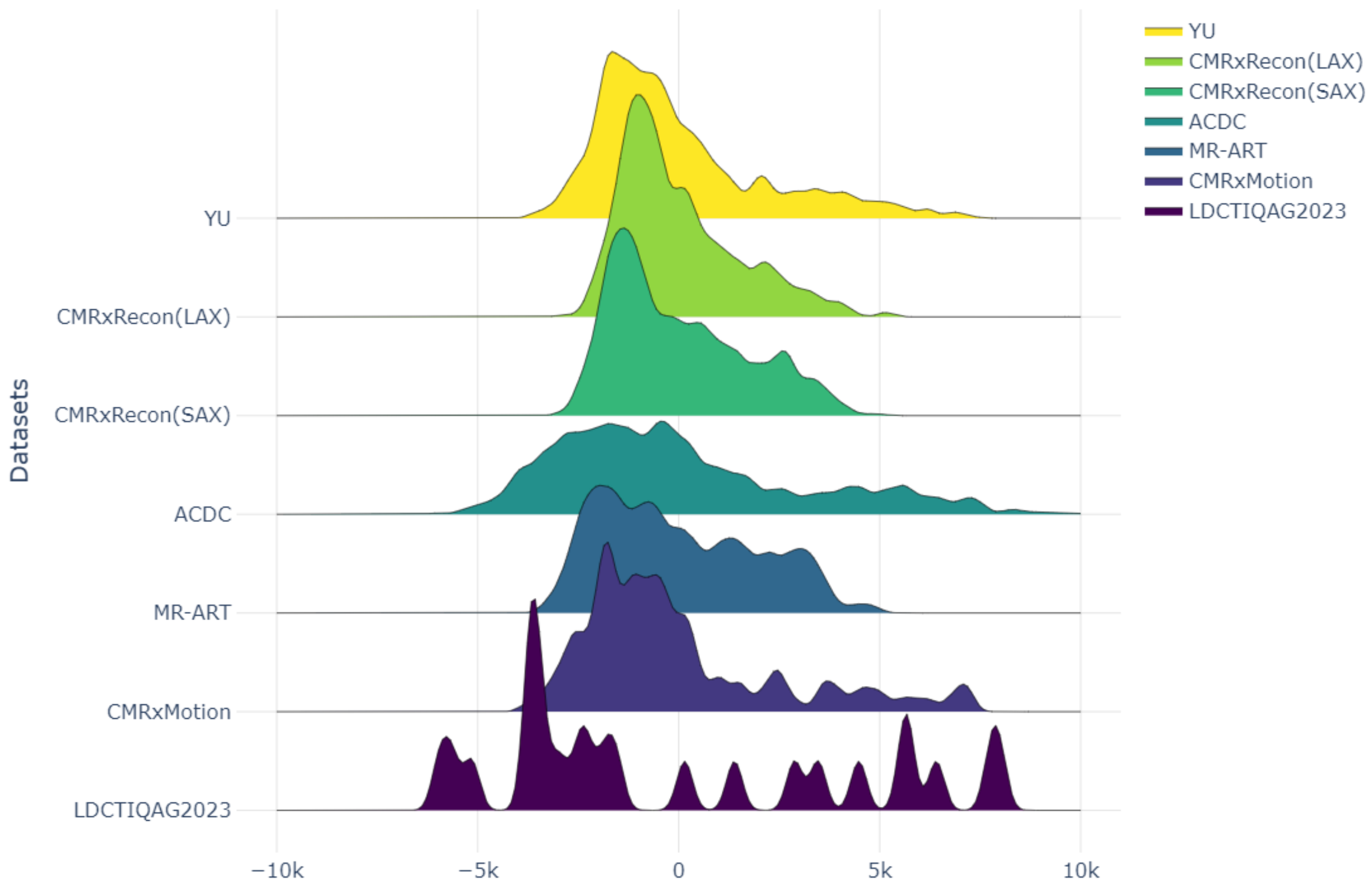

**Figure 2: The distribution of samples in the studied datasets.**

Table 3 conveys the results of using the proposed model to achieve a meticulous student model on the ACDC dataset. After training and testing the meticulous student model on the ACDC dataset as the target dataset, it has also been tested on the source datasets. The conducted experiments confirm the remarkable improvement of the results for the ACDC dataset after using the proposed model, while the performance is maintained for other source datasets. The experiment has been repeated in two modes by considering 32 (16 samples from each class) and 64 (32 samples from each class) random samples of test sets as an annotated set $\mathcal{D}_S^{Labelled}$. A small increase in the number of annotated samples also results in a remarkable growth in the results related to the target dataset, while the decrease in performance for the source datasets is negligible.

**Table 3: The results of the proposed model on the ACDC dataset and the source datasets. $\left|\mathcal{D}_S^{Labelled}\right|$ indicates the cardinality of this set.**

| | $\left\|\mathcal{D}_S^{Labelled}\right\| = 32$ | | | | $\left\|\mathcal{D}_S^{Labelled}\right\| = 64$ | | | |
|---|---|---|---|---|---|---|---|---|
| | ACC | PR | RE | F1 | ACC | PR | RE | F1 |
| ACDC | 89.40 ± 1.54 | 89.83 ± 1.45 | 89.07 ± 3.04 | 89.42 ± 1.88 | 95.85 ± 1.01 | 99.05 ± 0.38 | 92.66 ± 2.04 | 95.74 ± 1.12 |
| YU | 98.60 ± 0.26 | 98.36 ± 0.75 | 98.90 ± 0.40 | 98.62 ± 0.21 | 96.40 ± 0.66 | 99.48 ± 0.32 | 93.38 ± 0.32 | 96.33 ± 0.68 |
| CMRxRecon(LAX) | 98.51 ± 0.57 | 100.0 ± 0.0 | 97.00 ± 1.04 | 98.47 ± 0.54 | 97.12 ± 0.34 | 100.0 ± 0.0 | 94.12 ± 0.87 | 96.97 ± 0.46 |
| CMRxRecon(SAX) | 98.20 ± 0.66 | 100.0 ± 0.0 | 96.45 ± 1.29 | 98.19 ± 0.67 | 97.45 ± 0.80 | 100.0 ± 0.0 | 94.96 ± 1.59 | 97.41 ± 0.83 |

The analysis of the effectiveness of teacher models on the ACDC dataset in the current experiment and after training the meticulous student network through the proposed model is visualised in Figure 3. Visualisation is done using t-SNE dimensionality reduction approach (Van der Maaten & Hinton, 2008) for the test sets with $\left|\mathcal{D}_S^{Labelled}\right| = 64$. This figure acknowledges that the proposed model could reduce the domain shift between the source and target datasets and lead to improved results.

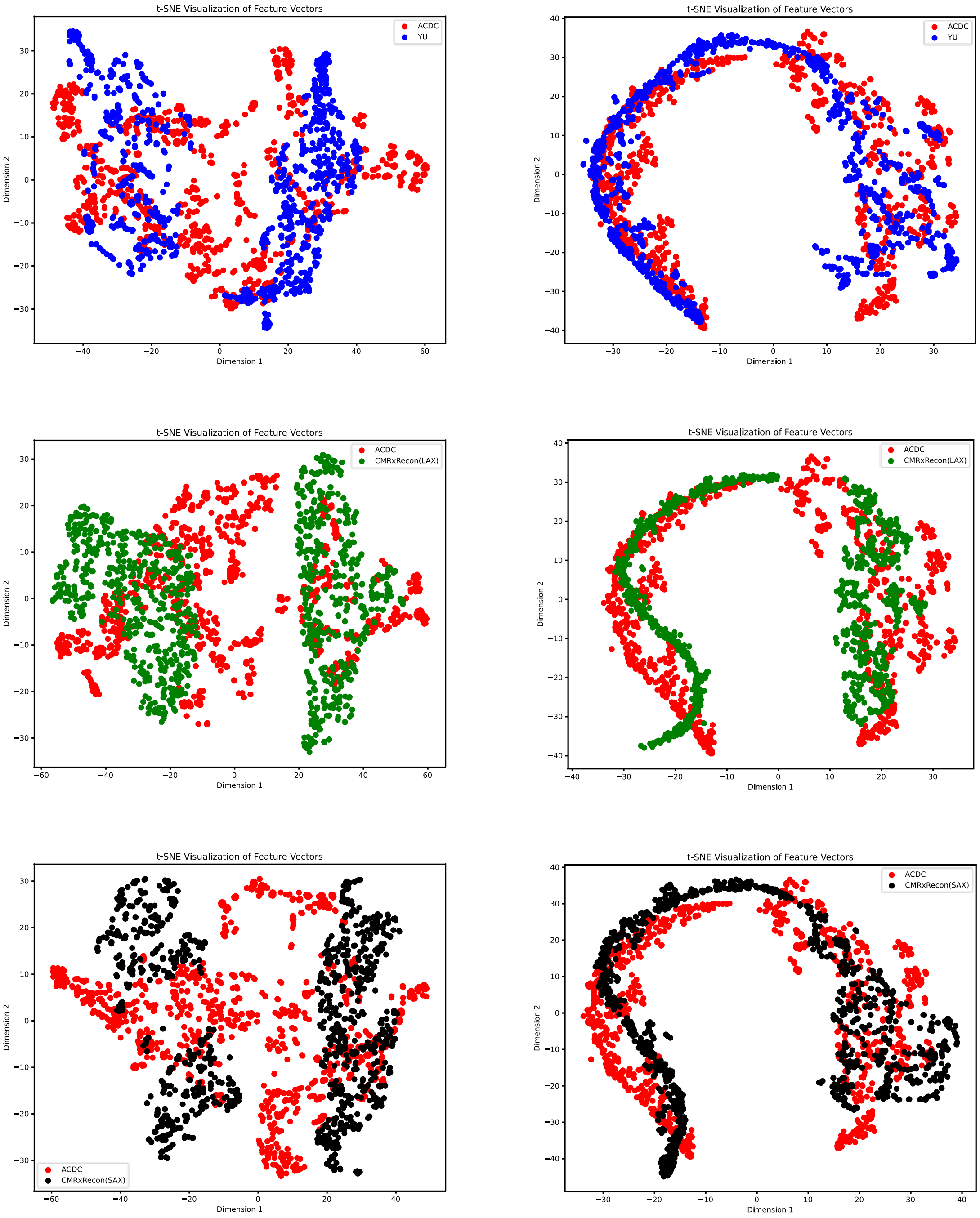

**Figure 3: Analysis of the effectiveness of teacher networks versus the meticulous student network trained through the proposed model. The results of visualizing test data on the teacher networks in the left column and the meticulous student network in the right column using the t-SNE approach.**

### 4-2- Cross Modality Analysis

Several experiments were conducted to train the meticulous student network on MR-ART and LDCTIQAG2023 datasets to evaluate the advantage of the proposed model on datasets with different modalities or contents than source ones. Although the different content of the images causes a considerable domain shift between the target and the source datasets in these experiments, the proposed model can take advantage of the common features of medical images to improve the efficiency of the meticulous student network. The results of these experiments are given in Table 4. As can be seen, the proposed model could improve the results, while as expected, it is understandable that due to the domain shift, the efficiency of the meticulous student network on the source datasets decreases. These experiments have been done to express the knowledge transfer capability of the proposed model on extremely dissimilar domains.

**Table 4: The results of the proposed model on the MR-ART and LDCTIQAG2023 datasets and the source datasets. $\left|\mathcal{D}_s^{Labelled}\right|$ indicates the cardinality of this set.**

| | $\left\|\mathcal{D}_S^{Labelled}\right\| = 32$ | | | | $\left\|\mathcal{D}_S^{Labelled}\right\| = 64$ | | | |
|---|---|---|---|---|---|---|---|---|
| | ACC | PR | RE | F1 | ACC | PR | RE | F1 |
| MR-ART | 82.22 ± 2.27 | 81.50 ± 3.02 | 82.08 ± 1.68 | 81.96 ± 2.32 | 91.20 ± 1.36 | 100 ± 0.00 | 82.22 ± 2.43 | 90.22 ± 1.46 |
| YU | 55.70 ± 1.98 | 52.20 ± 1.49 | 82.25 ± 1.97 | 63.86 ± 1.22 | 60.10 ± 2.03 | 55.06 ± 1.71 | 99.70 ± 0.39 | 69.20 ± 1.47 |
| CMRxRecon(LAX) | 61.00 ± 0.89 | 56.46 ± 0.98 | 98.39 ± 1.28 | 71.75 ± 1.15 | 89.63 ± 1.02 | 81.56 ± 1.87 | 100 ± 0.00 | 89.92 ± 0.00 |
| CMRxRecon(SAX) | 64.15 ± 1.09 | 91.02 ± 2.75 | 32.43 ± 2.62 | 47.47 ± 2.08 | 88.10 ± 0.98 | 81.16 ± 1.29 | 99.59 ± 0.38 | 89.43 ± 0.91 |
| | | | | | | | | |
| LDCTIQAG2023 | 74.00 ± 8.48 | 82.48 ± 16.62 | 70.62 ± 8.44 | 74.99 ± 8.50 | 86.89 ± 10.90 | 80.44 ± 15.68 | 100.00 ± 0.00 | 88.12 ± 11.35 |
| YU | 50.60 ± 1.30 | 50.60 ± 1.30 | 100.00 ± 0.00 | 67.19 ± 1.15 | 49.40 ± 1.54 | 51.35 ± 1.30 | 100.00 ± 0.00 | 68.19 ± 1.15 |
| CMRxRecon(LAX) | 49.55 ± 2.06 | 49.35 ± 2.02 | 100.00 ± 0.00 | 66.07 ± 1.80 | 50.85 ± 2.36 | 50.65 ± 2.02 | 100.00 ± 0.00 | 67.27 ± 1.80 |
| CMRxRecon(SAX) | 51.00 ± 1.24 | 50.80 ± 1.26 | 100.00 ± 0.00 | 67.37 ± 1.11 | 49.40 ± 1.16 | 49.940 ± 1.26 | 100.00 ± 0.00 | 65.25 ± 1.11 |

## 4-3- Evaluations on Real Artefact

Since the source datasets used to train teachers for identifying the respiratory motion artefact have been artificially degraded using k-space manipulation, in this experiment, the CMRxMotion dataset, which has real respiratory artefact data, was used to verify the validity of the proposed model. Table 5 presents the results of this experiment. Considering that the size of the CMRxMotion dataset is smaller compared to the source datasets (the size of the CMRxMotion dataset is 0.15 of the size of each of the source ones), it is expected that the performance of the proposed model shows less improvement than its experiment on the ACDC dataset. However, the trend of improving the results in this experiment is the same as the experiment on the ACDC dataset.

**Table 5: The results of the proposed model on the CMRxMotion dataset and the source datasets. $\left|\mathcal{D}_S^{Labelled}\right|$ indicates the cardinality of this set.**

| | $\left\|\mathcal{D}_S^{Labelled}\right\| = 32$ | | | | $\left\|\mathcal{D}_S^{Labelled}\right\| = 64$ | | | |
|---|---|---|---|---|---|---|---|---|
| | ACC | PR | RE | F1 | ACC | PR | RE | F1 |
| CMRxMotion | 66.15 ± 1.97 | 60.27 ± 1.95 | 99.60 ± 0.36 | 74.30 ± 1.52 | 71.43 ± 1.68 | 64.43 ± 1.69 | 97.74 ± 1.10 | 77.58 ± 1.42 |
| YU | 62.50 ± 3.11 | 57.59 ± 2.52 | 96.22 ± 1.74 | 72.19 ± 2.38 | 83.94 ± 1.19 | 77.94 ± 1.69 | 95.16 ± 1.21 | 87.12 ± 1.07 |
| CMRxRecon(LAX) | 62.63 ± 1.73 | 57.32 ± 2.09 | 92.61 ± 1.99 | 70.79 ± 1.92 | 85.61 ± 0.19 | 93.43 ± 0.78 | 76.08 ± 0.87 | 83.87 ± 0.33 |
| CMRxRecon(SAX) | 61.57 ± 4.71 | 57.83 ± 5.88 | 92.86 ± 2.38 | 71.10 ± 4.69 | 87.99 ± 1.01 | 85.12 ± 1.94 | 92.49 ± 0.31 | 88.64 ± 0.91 |

## 4-4- The Proposed Method vs the Related Studies

To make a fair comparison with previous related studies, the experiment of using YU, CMRxRecon(LAX) and CMRxRecon(SAX) as the source datasets and ACDC as the target with $\left|\mathcal{D}_S^{Labelled}\right| = 64$ was done. The results of these comparisons are shown in Table 6.

**Table 6: Comparison of the proposed method versus the related studies.**

| Model \ Metric | Accuracy | Precision | Recall | F1-score |
|---|---|---|---|---|
| Unsupervised Domain Adaptation based on Knowledge Distillation (Belal et al., 2021) | 82.43 ± 1.88 | 88.46 ± 1.33 | 72.67 ± 2.89 | 79.37 ± 1.89 |
| Unsupervised Domain Adaptation (Nabavi, Moghaddam, et al., 2023) | 82.73 ± 3.53 | 88.75 ± 1.20 | 79.41 ± 2.73 | 83.44 ± 2.29 |
| Transfer Learning (Tan et al., 2018) | 83.00 ± 1.75 | 91.85 ± 1.64 | 73.54 ± 3.66 | 81.62 ± 2.10 |
| Optimisation Trajectory Distillation (J. Fan et al., 2023) | 84.39 ± 2.03 | 93.08 ± 1.99 | 75.33 ± 3.19 | 83.68 ± 1.93 |
| Contrastive Knowledge Distillation (Xing et al., 2021) | 88.93 ± 1.54 | 95.38 ± 1.77 | 81.43 ± 2.65 | 88.44 ± 1.63 |
| Meta-Learning (Nabavi, Simchi, et al., 2023) | 89.67 ± 1.36 | 96.69 ± 1.18 | 83.58 ± 2.38 | 90.58 ± 1.45 |
| **The Proposed Model** | **95.85 ± 1.01** | **99.05 ± 0.38** | **92.66 ± 2.04** | **95.74 ± 1.12** |

## 4-5- Ablation Studies

Ablation studies were performed to evaluate the performance changes of the proposed model regarding the removal of influential elements, including the teacher $T_{DA}$, to extract domain-independent features and the filter $F$ to more accurately aggregate the feature vectors obtained from teachers. The results of these investigations are given in Table 7.

**Table 7: The results of ablation studies regarding the removal of influential elements, including $T_{DA}$ and $F$, to test the meticulous student model on the ACDC dataset as a target and the source datasets. $|\mathcal{D}_s^{Labelled}|$ is 64 and the numbers in parentheses indicate the standard deviation.**

| | ACDC | | | | YU | | | | CMRxRecon(LAX) | | | | CMRxRecon(SAX) | | | |
|---|---|---|---|---|---|---|---|---|---|---|---|---|---|---|---|---|
| | ACC | PR | RE | F1 | ACC | PR | RE | F1 | ACC | PR | RE | F1 | ACC | PR | RE | F1 |
| The Proposed Model (Excluding $T_{DA}$ and $F$) | 79.30 (1.84) | 97.90 (0.82) | 60.33 (4.05) | 74.58 (3.13) | 87.05 (0.83) | 89.40 (1.50) | 88.93 (0.38) | 89.16 (0.69) | 65.82 (1.67) | 99.38 (0.76) | 30.62 (3.51) | 46.69 (4.12) | 75.20 (1.03) | 100.0 (0.0) | 50.94 (2.65) | 67.46 (2.35) |
| The Proposed Model (Excluding $T_{DA}$) | 82.30 (1.21) | 98.23 (0.42) | 66.13 (3.09) | 79.01 (2.32) | 93.35 (0.97) | 92.74 (1.00) | 93.38 (0.32) | 93.50 (0.83) | 94.44 (0.90) | 98.23 (1.28) | 90.29 (1.77) | 94.08 (1.11) | 94.80 (1.58) | 97.70 (1.08) | 91.87 (2.62) | 94.68 (1.68) |
| The Proposed Model (Excluding $F$) | 83.80 (1.50) | 98.60 (0.74) | 68.98 (2.70) | 81.14 (1.89) | 89.40 (0.86) | 85.48 (0.88) | 94.27 (1.10) | 90.10 (0.66) | 88.98 (1.25) | 99.87 (0.27) | 77.63 (3.04) | 87.32 (1.93) | 91.45 (1.09) | 99.89 (0.22) | 83.16 (2.52) | 90.74 (1.47) |
| **The Proposed Model** | **95.85 (1.01)** | **99.05 (0.38)** | **92.66 (2.04)** | **95.74 (1.12)** | **96.40 (0.66)** | **99.48 (0.32)** | **95.26 (0.48)** | **96.33 (0.68)** | **97.12 (0.34)** | **100.0 (0.0)** | **94.12 (0.87)** | **96.97 (0.46)** | **97.45 (0.80)** | **100.0 (0.0)** | **94.96 (1.59)** | **97.41 (0.83)** |

Ablation studies were also conducted to investigate the effect of changing the number of teachers in the proposed model. To achieve this goal, the number of teachers was reduced to three, and the proposed model was tested with different teachers in two various experiments. The results of these experiments are tabulated in Table 8.

**Table 8: The results of ablation studies related to changing the number of teachers, to test the meticulous student model on the ACDC dataset as a target and the source datasets. $|\mathcal{D}_s^{Labelled}|$ is 64 and the numbers in parentheses indicate the standard deviation.**

| | ACDC | | | | YU | | | | CMRxRecon(LAX) | | | | CMRxRecon(SAX) | | | |
|---|---|---|---|---|---|---|---|---|---|---|---|---|---|---|---|---|
| Teachers | ACC | PR | RE | F1 | ACC | PR | RE | F1 | ACC | PR | RE | F1 | ACC | PR | RE | F1 |
| $T_{CMRxRecon(LAX)}$ $T_{CMRxRecon(SAX)}$ $T_{DA}$ | 94.32 (0.83) | 98.14 (0.78) | 91.57 (1.09) | 94.69 (0.87) | 92.00 (0.92) | 99.03 (1.03) | 85.03 (2.93) | 91.45 (1.27) | 97.97 (0.58) | 100.0 (0.0) | 95.88 (1.07) | 97.90 (0.56) | 98.20 (0.71) | 100.0 (0.0) | 96.44 (1.41) | 98.19 (0.73) |
| $T_{YU}$<br>$T_{CMRxRecon(SAX)}$<br>$T_{DA}$ | 91.65 (1.16) | 88.31 (2.03) | 96.24 (0.73) | 92.09 (1.16) | 97.5 (0.74) | 99.9 (0.36) | 95.60 (0.15) | 97.49 (0.76) | 94.00 (1.25) | 89.74 (2.07) | 99.61 (0.37) | 94.40 (1.08) | 97.50 (0.74) | 99.10 (0.36) | 95.95 (1.49) | 97.48 (0.76) |
| $T_{YU}$<br>$T_{CMRxRecon(LAX)}$<br>$T_{CMRxRecon(SAX)}$<br>$T_{DA}$ | 95.85 (1.01) | 99.05 (0.38) | 92.66 (2.04) | 95.74 (1.12) | 96.40 (0.66) | 99.48 (0.32) | 95.26 (0.48) | 96.33 (0.68) | 97.12 (0.34) | 100.0 (0.0) | 94.12 (0.87) | 96.97 (0.46) | 97.45 (0.80) | 100.0 (0.0) | 94.96 (1.59) | 97.41 (0.83) |

One of the hyperparameters that has a major impact on the performance of the proposed model is the number of epochs required for prior training and fine-tuning the meticulous student model. An ablation study was performed to achieve the appropriate trade-off between the number of epochs in these two steps. The results of this ablation study are shown in Figure 4.

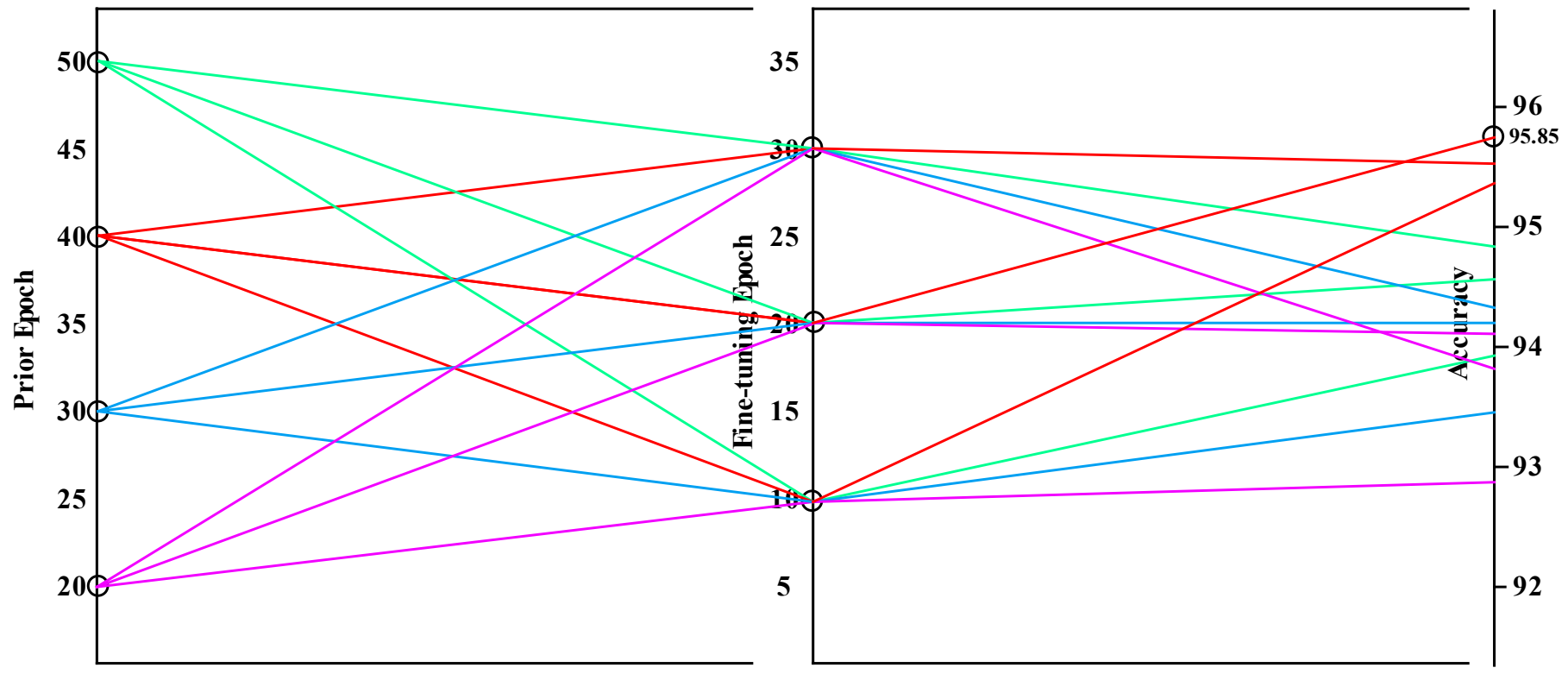

**Figure 4: The ablation study to adjust the number of epochs in the prior training and fine-tuning steps in the meticulous student network learning.**

## 5- Discussion

This study presents a new strategy for medical image classification that seeks to achieve multiple objectives simultaneously. These objectives include the adaptability of the model in the face of unseen data from various distributions, using knowledge transfer to train the model when there is a lack of access to bulk annotated data and the ability to privacy preservation by avoiding the public release of patient data. Besides, this model can also consolidate the tasks performed by several deep neural networks into one with almost similar performance, which facilitates the possibility of using it on various platforms. Accordingly, a novel domain adaptive strategy based on knowledge distillation, not a neural network architecture or deep learning model, was proposed, and its effectiveness was proved by extensive experiments. Thus, the performance of the introduced strategy will not be limited over time with the introduction and development of new deep learning architectures, and the networks used in this study can be replaced by new architectures.

By receiving the parameters of the trained teacher networks and the proposed knowledge distillation process, the proposed model trains a new network with a small number of annotated data from the target dataset, which can perform well on the target and also source data simultaneously. Although deep networks may perform well on trained data, their performance drops when faced with unseen data. The proposed model can guarantee the proper performance of the model on unseen data with different distributions due to the domain adaptability. Besides, the lack of access to data labels in medical applications has always been a serious challenge in training learning models. Due to their data-hungry nature, deep networks require large amounts of labelled data, while labelling medical data requires a costly and tedious process. The proposed model also addresses this challenge, and with a small amount of annotated data, it can significantly improve the final performance.

The proposed model was faced with data from different modalities and contents compared to what teacher networks had learned. In these experiments, the ability of the model to learn with very little labelled data was revealed. The issue that can be seen in the results of these experiments is the accuracy drop of the meticulous student network on the source data. In explaining this issue, it should be noted that teacher networks are trained on CMR images, while the target data are brain MRI and CT images, which have very different distributions compared to the source data. The knowledge distillatory section of the proposed model has been able to provide appropriate knowledge for training the meticulous student network, but in the fine-tuning stage, due to the striking difference between the source and target data, the knowledge of the source data has undergone changes. Based on the other experiments, we can be sure that by training teacher networks on data with the same modality and content as the target dataset, the proper performance of the meticulous student network can be achieved on the source and target data simultaneously.

We evaluated the proposed model for the task of respiratory motion artefact detection. Accordingly, synthetic but realistic artefacts were added to the imaging data of some datasets for this purpose. Thus, we also arranged an experiment to prove the model's performance on data with real artefacts. The investigations show a similar trend for the performance of the proposed model on the dataset with real artefacts. However, the smaller dataset is the reason for the lower metrics obtained in Table 5 compared to Table 3. The size of the CMRxMotion dataset is 0.15 times smaller than the ACDC dataset used in this study.

Comparing the proposed model with related studies shows its better performance. In these comparisons, related methods that were more aligned with the objectives of the current study were examined. Besides, the proposed model was evaluated from the point of view of the different parts used in it and the hyperparameters used in the experiments. Having a teacher network to extract domain-independent features and also using an embedding filtering process resulted in increasing the performance of the student network. The embedding filtering with the help of D and using domain-independent features with a remarkable impact on the efficiency of the student network led us to choose the meticulous student title. Moreover, the analysis shows that increasing the variety of data distributions in the form of growing the number of teachers results in improving the efficiency of the student network on the target dataset. However, this increase in diversity can mitigate the focus of the knowledge distillation process on a specific teacher, which reduces the efficiency of the student network on that dataset. As a result, the reduction of diversity in the process of knowledge distillation increases the efficiency of the student network on the dataset where the teacher trained on that dataset is available among the teacher networks.

## 6- Conclusion

This study presents a domain adaptive model based on knowledge distillation for medical image classification. This model can classify medical images in the presence of domain shifts between datasets and the absence of bulk annotated datasets. The proposed model does not require access to training data for developing teacher networks, so it improves the patients' privacy, and the classification process can only be carried out with the parameters of the trained teacher networks. Besides, the proposed model can consolidate the knowledge of multiple teacher networks into a meticulous student network with equivalent network architecture and almost similar performance. Hence, the proposed model as a strategy for medical image classification can address the fundamental challenges in the field simultaneously. Based on investigations, the proposed model outperforms previous studies. In the future study, we intend to investigate the efficiency of the proposed model for multi-class classification problems.

**Conflicts of Interests statement:** There are no conflicts of interests to declare.

**Data availability statement:** The datasets are available at

- York University (YU): http://jtl.lassonde.yorku.ca/software/datasets/
- CMRxRecon: https://cmrxrecon.github.io/
- Automated Cardiac Diagnosis Challenge (ACDC): https://www.creatis.insa-lyon.fr/Challenge/acdc/databases.html
- CMRxMotion: http://cmr.miccai.cloud/
- Movement-related artefacts (MR-ART): https://openneuro.org/datasets/ds004173/versions/1.0.2
- Low-dose Computed Tomography Perceptual Image Quality Assessment Grand Challenge (LDCTIQAG2023): https://zenodo.org/records/7833096#.ZEFywOxBzn5

**Code availability statement:** The codes are available at

https://github.com/kiananvari/MTMS-A-Domain-Adaptive-Meta-Knowledge-Distillation-Model-for-Medical-Image-Classification